\documentclass[twocolumn,showpacs,preprintnumbers,amsmath,amssymb,prl,superscriptaddress]{revtex4}

\usepackage{graphicx}
\usepackage{dcolumn}
\usepackage{bm}

\newcommand{\be}{\begin{equation}}
\newcommand{\ee}{\end{equation}}

\newcommand{\beq}{\begin{eqnarray}}
\newcommand{\eeq}{\end{eqnarray}}

\def\eq#1{(\ref{#1})}
\def\H1{\widehat{H}_1}

\begin{document}

\title{Competing effects of interactions and spin-orbit coupling in a 
quantum wire}

\author{V. Gritsev}
\affiliation{D\'epartement de Physique, Universit\'e de Fribourg,
CH-1700 Fribourg, Switzerland}

\author{G. I. Japaridze}
\affiliation{Institute of Physics, Georgian Academy of Sciences, Tamarashvili 6, 0177 Tbilisi, Georgia}

\author{M. Pletyukhov}
\affiliation{Institut f\"ur Theoretische Festk\"orperphysik, Universit\"at
Karlsruhe, D-76128 Karlsruhe, Germany}

\author{D. Baeriswyl}
\affiliation{D\'epartement de Physique, Universit\'e de Fribourg,
CH-1700 Fribourg, Switzerland}

\begin{abstract}
We study the interplay of electron-electron interactions and Rashba
spin-orbit coupling in one-dimensional ballistic wires. Using the
renorma\-lization group  approach we construct the phase diagram
in terms of Rashba coupling, Tomonaga-Luttinger stiffness
and backward scatte\-ring strength. We identify the parameter regimes
with a dynamically generated spin gap and show where the Luttinger
liquid prevails.
We also discuss the consequences for the operation of the
Datta-Das transistor.
\end{abstract}
\maketitle
The main goal of recent studies in the field of spintronics is to 
invent ways of manipulating the electron spin with an efficiency
comparable to that of present-day electronics (which manipulates 
charge) \cite{ZFS}.
One of the elementary spintronic devices, the Datta-Das transistor \cite{DD},
has been proposed more than a decade ago, its basic ingredient being a
ballistic
quantum wire with sufficiently strong Rashba spin-orbit interaction
\cite{Rashba}. The latter is required for creating a sizeable spin precession.
Depending on spin orientations in the source and in the drain one can
modulate the current flowing through the device and thus implement in 
principle
ON/OFF states.
The strength of the spin-orbit coupling can be tuned by applying a 
gate voltage to the
system \cite{Nitta}.

To understand the feasibility of such a device as well as its basic operation
it is important to investigate the effects of spin-orbit coupling on both
transport and magnetic properties of (essentially one-dimensional)
interac\-ting
electrons. Recently some progress towards a solution of this problem 
has been made
on the basis of the Tomonaga-Luttinger model \cite{MSB}.
In the present article we show that the {\it a priori}
assumption of validity of this approximation
is not always justified. We find in fact that the combined effects of
Coulomb interaction and Rashba coupling can generate a spin gap and thus
radically change the physical characteristics of the Datta-Das transistor. In
particular, we establish the parameter range  where the correlation
functions decay exponentially, and the Datta-Das device
becomes non-operating.

We consider a narrow ballistic wire described by a  one-dimensional
model of interacting electrons, including the Rashba-type spin-orbit coupling, 
$V_R=\alpha_R \sigma_y p_x$, where $\alpha_R$ is the coupling
strength and $\sigma_y$ is a Pauli matrix.

The electron-electron interaction (assumed to be weak) may be
decomposed into the different scattering  processes \cite{Gia}. We include
not only the standard forward scatterings of the
Tomonaga-Luttinger model (coupling parameters
$g_{1\parallel}, g_{2\parallel},g_{2\perp}$, $g_{4\parallel}$, $g_{4\perp}$),
but also the backward scattering term ($g_{1\perp}$). At the same 
time we do 
neglect Umklapp processes ($g_{3}$) since the system 
considered here is far 
away from half-filling.

Although in the 
absence of spin-orbit coupling (and for repulsive 
electron-electron
interactions) the backscattering term is usually
irrelevant, this is not always the case in the presence of spin-orbit 
coupling. As will be shown below, the combined effects of these two 
processes together with the forward scattering 
lead to the dynamical generation of a spin gap in the 
excitation spectrum in a wide range of 
coupling constants. In 
this range the ground state phase diagram qualitatively differs
from that of the Tomonaga-Luttinger model used before \cite{MSB,Haus}.

We assume weak bare couplings and therefore
linearize the single-particle spectrum around the two Fermi points
$\pm k_{F}$.  Decomposing the field operators in the standard way \cite{Gia}
into right ($r=+$) and left movers ($r=-$),
$\psi_{\sigma} (x) =\sum_r\psi_{r,\sigma} (x) \exp(irk_{F}x)$,
we obtain the con\-ti\-nu\-um limit of the fermionic Ha\-mil\-to\-ni\-an
$H=H_0+H_{R}+H_{\text{int}}$, where $H_0$ and $H_{\text{int}}$ are the usual one- and
two-particle parts, respectively, while the spin-orbit term reads
\begin{equation}
\!H_{R}\!=\!i\alpha_{R}k_F\!\!\sum_{r=\pm}\!r\!\!\int\!dx
(\psi^{\dag}_{r,\!\uparrow}(x)\psi_{r,\!\downarrow}(x)\!-
\!\psi^{\dag}_{r,\!\downarrow}(x)\psi_{r,\!\uparrow}(x)).
\label{Comtinuum}
\vspace{-0.1cm}
\end{equation}
We remark that terms involving gradients of the fields $\psi_{r,\sigma}(x)$
have been neglected because they would generate irrelevant operators
for non-vanishing electron-electron interactions \cite{remark}.
All parts of the Hamiltonian can be bosonized in terms of fields
$\phi_c(x), \theta_c(x)$ for the charge and $\phi_s(x), \theta_s(x)$
for the spin degrees of freedom, with commutation relations
$[\phi_{\lambda}(x),\theta_{\lambda'}(x')]\!=
\!(-i/2) \delta_{\lambda\lambda'} {\rm sign}(x-x')$ \cite{Gia}.
The Rashba term
\begin{equation}
H_{R} =  \frac{4\alpha_{R}k_F}{\pi \alpha}\!\int \! dx
\sin(\sqrt{2\pi}\phi_{s})\sin(\sqrt{2\pi}\theta_{s})\
\label{Bosonized}
\vspace{-0.1cm}
\end{equation}
(with a short distance cut-off $\alpha$) involves only $\phi_s$ and $\theta_s$ and
therefore does not affect
spin-charge separation (the same holds for the backscattering term). 
The kinetic energy and the forward scattering terms become $H=H_{c,0}+H_{s,0}$,
where the charge part $H_{c,0}$ and the free spin part $H_{s,0}$ both 
have the familiar
Tomonaga-Luttinger form,
\begin{equation}
H_{\lambda,0} = \frac{u_{\lambda}}{2}\int dx\ [K_{\lambda} 
(\partial_{x}\theta_{\lambda})^{2}
+\frac{1}{K_{\lambda}} (\partial_{x}\phi_{\lambda})^{2}]\ .
\label{Charge}
\end{equation}
The parameters $u_{\lambda}, K_{\lambda}$,
$\lambda=c,s$ depend in a simple way on
the coupling constants
$g_{1,\parallel}, g_{2,\parallel}, g_{2,\perp}, g_{4,\parallel},g_{4,\perp}$ \cite{Gia, GNT}.

Treating $H_R$ as a perturbation we notice that its vacuum expectation value
vanishes (a consequence of non-zero {\it conformal spin}).
Thus the lowest-order effect of the Rashba one-particle
process is seen to be absent, even though the process is strongly relevant.
One then has to study the higher-order expansions in order to find
contributions  with zero conformal spin (cf.\ Ref.\ \cite{Yakov,GNT}).
The second-order contribution (proportional to $\!\alpha_{R}^{2}$)
corresponds to effective two-particle interactions
\beq\label{bf}
\psi_{+,\uparrow}^{\dag}\psi_{-,\uparrow}
\psi_{-,\downarrow}^{\dag}\psi_{+,\downarrow}+
\psi_{+,\uparrow}^{\dag}\psi_{+,\downarrow}\psi_{-,\uparrow}^{\dag}
\psi_{-,\downarrow}+h.c.\ ,
\eeq
a first term describing a backscattering process already included in
the Hamiltonian and a second term representing a spin-nonconserving process
which has not been met before.
We therefore have to add the new interaction term in the renormalization
procedure, with a vanishing initial coupling constant.
In this way the operator product expansion governing the renormalization group (RG) flow is closed.
We also notice that
the process in question usually emerges in the field-theoretical
description of systems with completely broken spin-rotational
symmetry \cite{GS,TS}.

The bosonized Hamiltonian for the spin part can be written as
$H_s=H_{s,0}+H_R+H_{\text{bf}}$, where $H_{s,0}$ and $H_R$ have been given
before and $H_{\text{bf}}$ includes the backscattering and spin flip parts, 
cf. Eq.~(\ref{bf}),
\begin{equation}
H_{\text{bf}}\!=\!\frac{2g_{1\perp}}{(2\pi\alpha)^{2}}
\!\int\!dx\cos(\sqrt{8\pi }\phi_{s})
+\frac{2g_{f}}{(2\pi\alpha)^{2}}\!\int\!dx\cos(\sqrt{8\pi}\theta_{s}).
\label{Interaction}
\end{equation}
The Hamiltonian $H_s$ has the dual symmetry
\be\label{duality}
\phi_{s}\longleftrightarrow \theta_{s}, \ \ \
K_{s}\longleftrightarrow K_{s}^{-1}, \ \ \
g_{1\perp}\longleftrightarrow  g_{f} ,
\ee
which will be used below to obtain the phase diagram
of the model in the full parameter range.
Thus, if a transition occurs for some set of parameters there must be also
a transition for the {\it dual} set of parameters.

Standard perturbative RG analysis \cite{Boyan} up to the
third order in couplings  yields the set of equations
\beq\label{1RG}
\frac{d y_{R}}{dl}&=& (4-(K_s + K_s^{-1}))y_{R}+y_{\perp}y_{f}y_{R},
\nonumber\\
\frac{dy_{\perp}}{dl}&=&2(1-K_s ) y_{\perp} +(K_s^{-1}-K_s )y_{R}+
\!\frac{y_{\perp}}{4}(y_{\perp}^{2}-y_{f}^{2}),
\nonumber\\
\frac{dy_{f}}{dl}&=& 2(1-K_s^{-1})y_{f} -(K_s^{-1}-K_s )y_{R}
-\!\frac{y_{f}}{4}(y_{\perp}^{2}-y_{f}^{2}),
\nonumber \\
\frac{d K_s}{dl}&=&\frac{1}{2} \left(y_{f}^{2} -y_{\perp}^{2} K_s^2 \right),
\eeq
where $l$ measures the logarithm of the length scale, 
$y_{\perp}= g_{1\perp}/\pi u_{s}$,
$y_{f}=g_{f}/\pi u_{s}$ and $y_R\!=\!(2\alpha_R k_{F}\alpha/\pi u_{s})^2$.
Note that it is $y_R$ that enters \eq{1RG}, not $\alpha_R$. The RG equations
are therefore independent of sign$(\alpha_R)$, a consequence of time-reversal
symmetry. The initial conditions are $K_s(0)\!\equiv\!K_{0}$,
$y_{\perp}(0)\!\equiv\!y_{\perp,0}$, $y_{R}(0)\!\equiv y_{R,0}$ and $y_{f}(0)\!=\!0$.
Similar RG equations have been derived in the context of spinless fermions on
a ladder \cite{Yakov, GNT}. The RG equations \eq{1RG} have three
weak-coupling fixed points: (I) $y^*_{\perp}=y^*_f=y^*_{R}=0$ and
$K^*_s$ arbitrary ; (IIa,b) $K^*_{s}=1$, $y^*_{R}=0$, $y^*_{f}=\pm y^*_{\perp}$
and $y^*_{\perp}$ arbitrary.
The spin-wave fixed point \cite{JKKN} (I) corresponds to the noninteracting
system with no spin-orbit coupling. The fixed points (IIa,b) correspond to the
critical Ashkin-Teller (AT) model \cite{JKKN,TS}.

A rough idea of the RG flow can be given by assuming the spin stiffness $K_s$ 
to be constant
(=$K_0$) and by neglecting cubic terms in (\ref{1RG}). The solutions are 
\beq\label{AS}
y_{R}(l)&=& y_{R,0} e^{l \gamma_{R}},\nonumber\\
y_{\perp,(f)}(l)&=& A_{\perp,(f)} e^{2 l [1-K_0^{(-1)}]}
+B_{\perp,(f)} e^{l \gamma_{R}}, 
\eeq 
where $B_{\perp,(f)}\!=\pm y_{R,0}[K_0^{-1} -K_0]/[2 \pm K_0 \mp K_0^{-1}]$ 
and $A_{\perp,(f)} = y_{\perp,(f),0} - B_{\perp,(f)}$. The one-particle Rashba
process is relevant if the exponent $\gamma_{R}=[4- (K_0 + K_0^{-1})]$ is 
positive, 
i.e. for $2-\sqrt{3}<K_{0}<2+\sqrt{3}$. The amplitude $y_{R,0}$ is larger than
both $B_{\perp}$ and $B_{f}$ for $(\sqrt{5}-1)/2<K_{0}<(\sqrt{5}+1)/2$. This
(Rashba dominated) region will be investigated more accurately below.
For $K_0\le (\sqrt{5}-1)/2$, backscattering ($y_{\perp}$) dominates, while for
$K_0\ge (\sqrt{5}+1)/2$, spin flip processes ($y_f$) prevail, in agreement
with the duality relations (\ref{duality}).

\begin{figure}[hb]
\vspace{-0.2cm}
\includegraphics[width =5.0cm,angle=270]{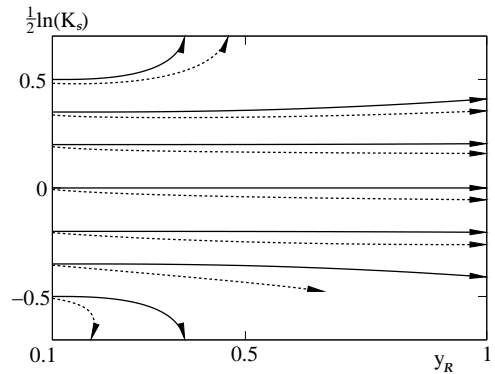}
\caption{Flow diagram for the $(y_R$,$\frac{1}{2}\ln K_s)$ 
plane for
$y_{R,0}=0.07$ and various initial values $K_{0}$. Full 
lines correspond to 
$y_{\perp,0}=0$, dashed lines to 
$y_{\perp,0}=0.4$. 
\label{flow}}
\end{figure}

We return now to the full RG equations (\ref{1RG}).
Numerical solutions for various initial conditions in the regime of dominant
spin-orbit coupling are illustrated in Fig.~\ref{flow} as flows in the
$\textstyle (y_{R},\frac{1}{2}\ln K_{s})$ plane. 
The results confirm 
that $K_{s}$
remains essentially constant for $y_{\perp,0}=0$ and is weakly renormalized
for $y_{\perp,0}=0.4$. On the other hand, the Rashba coupling $y_{R}$ rapidly
grows from its initial value $y_{R,0}$ up to some value of order 1 at a length
$l_{R}$, while the couplings $y_{\perp}$ and $y_{f}$ remain small.
It is then possible to add a second step in the RG procedure \cite{GNT}
by treating $H_R$ at $l_R$ non-perturbatively through
the canonical transformation
$\psi_{r,\pm} ={\textstyle \frac{1}{\sqrt{2}}}
\left( \psi_{r,\uparrow\downarrow} - i  \psi_{r,\downarrow\uparrow}\right)$.
The transformed single-particle Hamiltonian is
\begin{equation*}
H_{0}+H_{R}=\!\!\sum_{r,r'=\pm}\!\int \!dx \ \psi^{\dag}_{r,r'}(x)
(rv_{F}\partial_x+r'{\alpha}_{R}k_{F})\psi_{r,r'}(x) .
\label{hsp}
\end{equation*}

Its spectrum consists of two subbands split horizontally by
$2{\alpha}_R k_F/v_F$, where $v_F=u_sK_s$ is
the renormalized Fermi velocity.
The canonical transformation leads to the same type of interaction
terms
as in the original Hamiltonian, with coupling constants
$\tilde{y}_{\perp,f}\approx K_s-1+(y_{\perp,f} - y_{f, \perp})/2$,
$\tilde{K}_s -1\approx (y_{\perp} +y_f)/2$ (for $| K_s - 1 | \ll 1$;
otherwise the relations are slightly more complicated).

Bosonization can be applied to the new Hamiltonian in terms of fields
$\tilde{\phi}_s$ and $\tilde{\theta}_s$. The Rashba term
leads to a contribution
$(\sqrt{2}/\pi)\alpha_{R}k_F\int dx\partial_x \tilde{\theta}_s$ which can
be absorbed in the single-particle part by the shift
\be\label{canon}
\Theta_s (x)  = \tilde{\theta}_s (x) +k_{R}x,
\quad  \Phi_s (x) = \tilde{\phi}_s (x),
\ee
where $k_{R}=\sqrt{2}\alpha_{R}k_F/(\pi u_{s}\tilde{K}_{s})$. 
The spin part of the new Hamiltonian reads
\beq\label{incom}
\tilde{H}_s &=& \frac{u_{s}}{2}
\int dx \left\{ \tilde{K}_{s}(\partial_{x}\Theta_s )^{2}+
\frac{1}{\tilde{K}_{s}}(\partial_{x}\Phi_s )^{2}\right.  \\
& &+ \left. \!\frac{\tilde{y}_{\perp}}
{\pi \alpha^{2}}\cos(\sqrt{8\pi}\Phi_{s})-
\frac{\tilde{y}_{f}}{\pi \alpha^{2}}\cos(\sqrt{8\pi}
\Theta_{s}- k_{R}x) \right\}. \nonumber
\eeq
For $\tilde{K}_{s}>1$ the ordering
in the $\Theta_s$ sector and the Rashba process are competing, but
since for $y_{R}\approx 1$ and $\tilde{y}_{f}\ll 1$ the last term is 
wiped out by strong oscillations,
the subsequent flow involves only the parameters $\tilde{K}_s$ and
$\tilde{y}_{\perp}$. The equations of the second RG step have the form of the
Kosterlitz-Thouless flow
\be\label{2RG}
\frac{d\tilde{K}_{s}}{dl} = -\frac{1}{2}\tilde{y}_{\perp}^{2},
\quad
\frac{d\tilde{y}_{\perp}}{dl} = 2(1-\tilde{K}_s)\tilde{y}_{\perp},
\ee 
with the initial conditions
$\tilde{K}_{s}(0)=1+\frac{1}{2}(y_{\perp}(l_{R})+y_{f}(l_{R}))$,
$\tilde{y}_{\perp}(0)=K_s(l_R)-1 +\frac{1}{2}(y_{\perp}(l_{R})-y_{f}(l_{R}))$.
Eqs.\ (\ref{2RG}) imply two distinct scenarios,
for $|\tilde{y}_{\perp}|> 2[\tilde{K}_{s}(0)-1]$ a flow to strong coupling
and for $|\tilde{y}_{\perp}|<2[\tilde{K}_{s}(0)-1]$ a flow to the weak-coupling
fixed line  $\tilde{y}_{\perp}^*=0$,
$\tilde{K}^{*}_{s}$ (non-universal renormalized value).
Clearly the line $y_{\perp}=y_f$, $K_s=1$ is critical for arbitrary $y_R$.
We find systematically the weak-coupling case (for $y_{\perp,0}\ge 0$). Thus,
if the flow is dominated by the Rashba term, the system is a Luttinger liquid,
with enhanced spin precession.

We discuss now the parameter regions where the Rashba term is
not dominant, and the RG flow is governed by Eqs.\ (\ref{1RG}). We find
numerically that for $K_s\neq 1$ either $y_{\perp}$ or $y_f$ are
renormalized towards
strong coupling, $y_{\perp}$ for $K_s<1$, $y_f$ for $K_s>1$, in agreement
with the approximate analytical solutions (\ref{AS}). Both cases scale to the
strong-coupling regime of the sine-Gordon model and therefore must have a spin 
gap $\Delta_s$ in the excitation spectrum. At the same time, the Rashba term and the 
band splitting are dynamically suppressed. This is an unexpected new result, 
because in the absence of the bare spin-orbit coupling there is no spin gap 
for $y_{\perp}\ge 0$.

Our findings can be interpreted in terms of two commensurate-incommen\-surate
transitions \cite{JNPT}. In the incommensurate phase the Rashba term 
dominates and the effective field
theory is equivalent to a Tomonaga-Luttinger model with non-universal spin stiffness 
and momentum shift (\ref{canon}). In the commensurate phases
 the spin excitations have a gap, produced by
backscattering and spin flip processes, respectively.
\begin{figure}[ht]
\includegraphics[width =8.0cm]{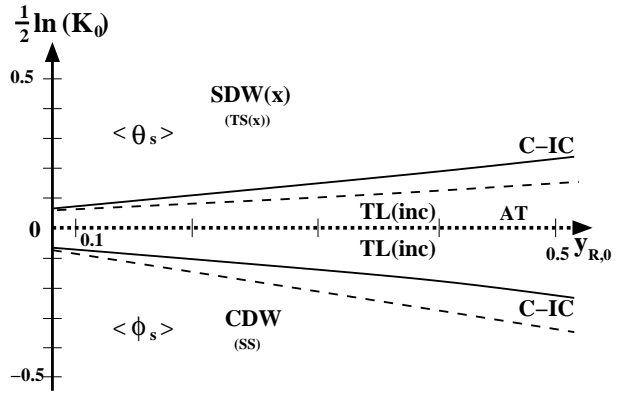}
\caption{Phase diagram in the parameter space of 
initial values $y_{R,0}$ and $\textstyle{\frac{1}{2}}\ln(K_{0})$ for $y_{\perp,0}=0$ 
(full lines), $y_{\perp,0}=0.4$ (dashed lines), and for $y_{f,0}$=0. The 
figure exhibits clearly the dual symmetry of Eq.\ (\ref{duality}). 
Spin gaps exist both in the phase with dominant superconducting 
correlations (SS or TS(x)) and in the spin (charge)-density wave phases 
(SDW(x) or CDW). In the Luttinger-liquid phase (TL(inc)) all correlation 
functions  decay algebraically. 
The TL phase is separated from the two spin-gapped phases by 
commensurate-incommensurate transitions (C-IC lines). The bold dotted line 
(AT at $K_{s}=1$) corresponds to the critical Ashkin-Teller model.} 
\label{g0}
\end{figure}
Fig.~\ref{g0} shows the phase diagram
in the parameter space $(y_{R,0} , \textstyle{\frac{1}{2}}\ln K_0)$ for the initial values
$y_{\perp,0}=0, 0.4$ and $y_{f,0}=0$. The straight line $\ln K_0=0$
(fixed points (II)) represents the self-dual line for $y_{\perp,0}=y_{f,0}=0$.
The region of the incommensurate Luttinger-liquid phase widens as a function
of $y_{R,0}$. Outside of this region, the mean values $\langle\phi_s\rangle$ 
or $\langle\theta_s\rangle$ are finite, the former for $K_0<1$, the latter for 
$K_0>1$. In the $\phi_s$-ordered phase the dominant correlations are
singlet superconductivity (SS, for $K_{c}>1$) and charge-density waves
(CDW, for $K_{c}<1$). In the $\theta_s$-ordered phase the dominant
correlations are the  $x$-component of spin-density waves (SDW$_{x}$,
for $K_{c}< 1$) and the $x$-component of triplet superconductivity
(TS$_{x}$, for $K_{c}>1$). 
The transition lines in Fig.~\ref{g0} have been determined approximately on 
the basis of the RG flow. We note that the exact characterization of these
commensurate-incommensurate transitions would require a non-perturbative 
analysis, which goes beyond the present perturbative RG scheme.

Spin precession is described by the correlation function
\be\label{fun}
f(x)=\textstyle{\frac{1}{2}}\langle (\psi_{\uparrow}(x)+
\psi_{\downarrow}(x))(\psi_{\uparrow}^{\dag}(0)
+\psi_{\downarrow}^{\dag}(0)) \rangle.
\ee
For a narrow, ballistic quantum
wire connecting a source at $x=0$ to a drain at $x=L$
the quantity $|f(L)|^{2}$ measures the probability for a particle
entering the drain to have the same spin orientation as one leaving the
source. Within the Tomonaga-Luttinger model
$f (x)$ is found \cite{Haus} to vary as
$|x|^{-\gamma}$, where $\gamma  =
(1/4)(K_{c}+1/K_{c}+K_{s}+1/K_{s})$ depends on the electron-electron
interaction
through the charge and spin stiffnesses
$K_c$ and $K_s$, respectively. The same behavior is
expected to occur in our case within the Luttinger-liquid phase except that
$K_s$ is replaced by the (non-universal) fixed-point value $K_s^*$.
In principle, the presence of irrelevant couplings $y_{\perp}$ and $y_{f}$ 
can lead to
extra multiplicative corrections \cite{Gia} to $f (x)$, but they appear to
vanish identically in our case.
In the spin-gapped regime where the ordering of the $\phi_{s}$ field
implies the disordering of $\theta_{s}$ field for $K_{s}<1$ (and vice versa
for $K_{s}>1$),
the function $f(x)$ is expected to decay as the corresponding correlator 
in the sine-Gordon model \cite{LZ}, namely $f(x)\sim x^{-(\lambda +1)}\exp(-x/\xi_s)$, 
where $\xi_s = \hbar v_F /\Delta_s$ is the correlation length, 
and $\lambda = (\sqrt{K_{c}}-1/\sqrt{K_{c}})^{2}/4$.

We note that the spin stiffness $K_{s}$ is an important parameter in the 
problem. In $SU(2)$-invariant models with repulsive interactions $K_{s}$ 
scales to the fixed-point value $K_{s}^{*}=1$. This is no longer true for 
systems with spin-orbit coupling where the spin-rotation symmetry is reduced 
to $U(1)$. 

As an example we consider a narrow InAs quantum wire with
$m^{*}=0.023m_{e}$ and $\alpha_{R}=(0.6-4)\times 10^{-11}$ eVm \cite{ZFS}.
We choose the wire width $d=5$ nm and the Fermi wave vector
$k_{F}=0.5\times 10^{8}$m$^{-1}$. In this case the assumption
of a single occupied subband is justified.
The parameter $y_{R,0}$ depends on 
the cut-off length $\alpha$, for which a natural choice is the width $d$. 
Using these values and $u_{s}\approx v_{F}$ we find that $y_{R,0}$ ranges 
from $10^{-4}$ to $10^{-2}$. In order to estimate $K_{0}$, we use the 
well-known relation  \cite{Gia} between the spin stiffness and the Fourier 
transform $V_{eff}(q)$ of the effective interaction potential for quantum 
wires. The latter is taken in the form proposed in Ref. \cite{hm}. Thus we obtain 
$\frac12 \textstyle{\ln K_0}\approx 0.15$. According to Fig.~\ref{g0} this 
corresponds to the spin-gapped phase. The standard procedure for the 
sine-Gordon model \cite{Gia} yields a value of  $\Delta_{s}$ in the range 
$(0.01-0.1)\varepsilon_{F}$, where $\varepsilon_{F}\approx 4$ meV for given 
$m^{*}$ and $k_{F}$. For temperatures $T\le\Delta_{s}/k_{B} = 0.5 - 5$ K 
and wire lengths exceeding the correlation length, $L \geq \xi_{s}=(0.4-4)\mu$m
the phenomena described above will play an important role. These regions 
can be reached in present-day devices ($L \approx 2-6 \mu$m), 
and it should in principle be possible to detect signatures of the spin gap.

If the material parameters can be tuned close to the 
commensurate-incommensurate
transition from a Luttinger 
liquid to a phase with a spin gap, the dependence
of the Rashba 
coupling on the electric field strength may allow to drive 
the
system from one side of the transition to the other. Such a 
control of a spin gap
by a gate voltage would represent a spectacular 
novel field effect. The
observation of such a subtle phenomenon would 
of course not only be of fundamental
interest, but it could also pave 
the way for the fabrication of new types of
devices.

In conclusion, we have found that the interplay between electron-electron
interactions and Rashba spin-orbit coupling in a narrow wire generates
a spin gap $\Delta_s$ for a certain range of parameters. This  restricts the
operation of the Datta-Das transistor.
In the Luttinger-liquid phase, where correlation functions fall off according
to power laws, the device may work, but in the spin-gapped phase the
spatial coherence of spin precession is suppressed exponentially, and the
device efficiency tends to zero at lengths
greater than the correlation length $\xi_s$.

We are grateful to Thierry Giamarchi and Gerd
Sch\"{o}n for valuable discussions. V.G. was supported by the Swiss
National Science Foundation through grant Nr.20-68047.02. G.I.J. and 
D.B. 
acknowledge support through the SCOPES grant Nr. 
7GEPJ62379.
M.P. was supported by the DFG Center for Functional 
Nanostructures at the 
University of Karlsruhe.


\begin{thebibliography}{99}
\bibitem{remark}
This can be checked within the perturbative scheme used in this paper.
\end{thebibliography}
\end{document}